\documentclass[pdflatex,sn-mathphys-ay]{sn-jnl}

\usepackage{graphicx}%
\usepackage{multirow}%
\usepackage{amsmath,amssymb,amsfonts}%
\usepackage{amsthm}%
\usepackage{mathrsfs}%
\usepackage[title]{appendix}%
\usepackage{xcolor}%
\usepackage{textcomp}%
\usepackage{manyfoot}%
\usepackage{booktabs}%
\usepackage{algorithm}%
\usepackage{algorithmicx}%
\usepackage{algpseudocode}%
\usepackage{listings}%
\usepackage{subcaption}%
\usepackage{float}
\usepackage[normalem]{ulem}

\renewcommand{\phi}{\varphi}

\renewcommand{\epsilon}{\varepsilon}
\newcommand{\celsius}{$^\circ$}
\newcommand{\unit}{$\mu\text{g}\cdot\text{m}^{-3}$}

\newcommand{\no}[1]{NO$_{#1}$}

\begin{document}

\title[Article Title]{Using low-cost sensors to improve NO$_2$ concentration maps derived from physico-chemical models}

\author*[1]{\fnm{Emma} \sur{Thulliez}}\email{emma.thulliez@insa-rouen.fr}

\author[2]{\fnm{Camille} \sur{Coron}}\email{camille.coron@inrae.fr}

\affil[1]{INSA Rouen Normandie, Normandie Univ, LMI UR 3226, F-76000 Rouen, France}

\affil[2]{Université Paris-Saclay, AgroParisTech, INRAE, UMR MIA Paris-Saclay, 91120, Palaiseau, France}


\abstract{Urban air quality is a major concern today. Concentrations of pollutants, such as \textcolor{black}{nitrogen dioxide}, must be monitored to ensure that they do not exceed hazardous thresholds. \textcolor{black}{For this reason, scarce reference stations, which are generally managed by air quality monitoring associations, are located in major cities.} Two recent approaches enable fine-scale mapping of pollutant concentrations. The first relies on deterministic physico-chemical models that incorporate the street network and compute concentration estimates on a grid, producing spatial maps. The second is based on the emergence of low-cost sensors, which enable monitoring organizations to increase the density of their measurement networks. However, these sensors are \textcolor{black}{unreliable} and require \textcolor{black}{regular and important} calibration. We propose \textcolor{black}{to combine these approaches and } improve maps generated by deterministic models by integrating  data from multiple sensor networks. Specifically, we model the bias of deterministic models and estimate its parameters using measurements, through a Bayesian nested framework. Our approach simultaneously enables the calibration of low-cost sensors and the correction of deterministic models outputs. This method, although general, is applied to the city of Rouen (France), combining outputs of the physico-chemical model SIRANE \textcolor{black}{\citep{SIRANE-1}} and the measurements provided both by 4 reference monitoring stations and 10 low-cost sensors during December 2022. Results show that the method indeed corrects the concentration maps, reducing the root mean squared error by approximately \textcolor{black}{12.4}\%, and that low-cost sensors play an essential role in this correction.}


\keywords{Air quality, Nitrogen dioxide, Sensor calibration, physico-chemical models, Bias modeling, Data fusion, AirSensEUR, SIRANE}



\maketitle

\section{Introduction}
Over the past few years, air quality has become a major concern in public policy. Among pollutants, nitrogen dioxide (\no2) is of particular importance. It is in particular known to exacerbate respiratory and cardiovascular condition in vulnerable populations such as children and the elderly (see \cite{FelberDifferences2008, WHO-review-NO2-2020, SaucyLUR2018, LeungShortTerm2021}). In response, the \cite{WHO_Standard_2021} has issued guidelines, and \no2 concentrations are now subject to legislation from major governmental entities, such as the European Union and the French government (\cite{seuils-no2}).

\no2 is primarily produced by traffic, and is therefore particularly abundant in urban areas (\cite{eea-no2}). To assess \no2 concentrations in cities, local stakeholders rely on two main sources of information: measurements and model outputs. Measurements provide real-time data, but are limited to a small number of locations. Model outputs, in contrast, enable the creation of fine-resolution maps, but depend on physico-chemical modeling that does not incorporate real-time information and may suffer from biases when assumptions about physical or meteorological conditions are not met. These two data sources are therefore very different yet complementary.

Air quality measurements are traditionally obtained from reference instruments (also known as monitoring stations) \textcolor{black}{that are generally operated by local air quality monitoring agencies. These instruments} are considered to be highly reliable \textcolor{black}{(i.e. are assumed to provide unbiased measures of the real concentration}\footnote{\color{black} in France, the list of instruments that lie in this category is regularly updated by the LCSQA (see \url{https://www.lcsqa.org}), according to the norm NF EN 14211 for nitrogen dioxide.}\textcolor{black}{)} but entail substantial costs. In recent years, the field of air quality monitoring has witnessed the emergence of new technologies designed to reduce these costs. In particular, low-cost sensors have become increasingly widespread. These devices are considerably less expensive (between 10 and 100 times cheaper than reference stations) and thus enable the deployment of denser monitoring networks. However, \textcolor{black}{low-cost sensors} may provide erroneous measurements under various conditions and therefore require calibration (\cite{EuNetAir_Joint_Exercise_2016,WeiImpact2018,ChristakisAir2023}). The standard calibration approach consists in collocating sensors with a reference station and comparing their measurements in order to build a calibration model. 
This procedure nonetheless presents important limitations: it does not account for temporal drifts of sensors behaviors and necessitates relocating the sensors after a certain period, during which they provide no additional information {beyond that of} the {reference} stations (see \cite{CastellKindergarten2018, LiCharacterizing2021}). Additionally, \cite{BobbiaStatistical2025} have shown that the behavior of low-cost sensors can change as a result of being moved, or even simply switched off and on again. 
Recent strategies have focused on calibrating sensors post-deployment using statistical or machine-learning techniques. For example, \cite{BobbiaSpatial2022} apply kriging methods, while \cite{WesselingBenchmark2024} propose clustering-based corrections.

Besides, several physico-chemical models are used to give estimations of pollutant concentrations in urban environments at any time and any location (see Section \ref{subsec:bias}). \textcolor{black}{Usually, they consist in general models that are applied by local organization to assess pollution at local scale. They generally combine physico-chemical reactions modeling and local data (like meteorological data) to  provide an estimation of the concentration at all locations of the area of interest.} However, these models inevitably suffer from bias --particularly at high \no2 concentrations-- because they cannot fully account for highly local and transient factors such as roadworks or real-time traffic \textcolor{black}{and generally do not integrate important information such that land use for instance}. Several methods exist for correcting model outputs. For example, data fusion techniques, such as kriging, integrate measurements into concentration maps (\cite{GressentDataFusion2020, ORegan-DataFusion-2024}). Another approach, based on multi-fidelity, consists in ordering data sources according to their reliability and cost (\cite{LeGratietBayesian2013, LeGratietRecursive2014}).

We introduce a new approach that combines physico-chemical model outputs and concentration measurements provided by both reference stations and low-cost sensors through a probabilistic model. This framework enables us to simultaneously calibrate low-cost sensors and correct physico-chemical model outputs. More specifically, we set a model for the bias of physico-chemical outputs and another for sensor measurements, and estimate both models parameters simultaneously. Our approach is general and these models must be adapted to the considered city and the sensors used. 
The proposed method enhances \no2\ mapping, improves the understanding of low-cost sensor behavior, and \textcolor{black}{in particular} enables the study of their temporal dynamics. This framework thus provides a method to combine multi-source measurements with deterministic modeling to produce more accurate air pollution maps. We notably find out that using micro-sensor measurements in addition to measurements from reference stations (and not only the latter) is essential to this improvement.

Our method is general and can be adapted to different urban contexts, different pollutants, different types of sensors, or different cities. In this article, we apply it to the estimation of \no2 concentrations in the city of Rouen, using SIRANE as physico-chemical model and combining its outputs to the \textcolor{black}{measurements} provided by four reference monitoring stations and ten low-cost sensors during December 2022. \textcolor{black}{For this specific situation we can provide a reduction by more than $12\%$ of the error of the concentration map. For more readability, we start our article by presenting this particular framework, but our approach aims at being adapted to each city.}

The paper is organized as follows. Section \ref{sec:data} describes the physico-chemical model and measurement data. Section \ref{sec:model} introduces the models for both physico-chemical output errors and sensor measurement calibration. Section \ref{sec:estim} presents the estimation procedure. Section \ref{sec:results} reports the results.

\section{Data}
\label{sec:data}
\subsection{Physico-chemical models outputs}

The physico-chemical model considered in this article is the computational fluid dynamics (CFD) model SIRANE \citep{SIRANE-1, SIRANE-2, SIRANE-3}. SIRANE simulates urban air pollution and is widely used by monitoring agencies to represent air quality at city scale with very high spatial resolution (one \no2 concentration estimate per 10 m $\times$ 10 m grid cell, per hour).
The model explicitly accounts for the road network, distinguishing between traffic and background areas. Moreover, it also incorporates meteorological parameters (e.g. ambient temperature, wind speed and direction, humidity), traffic estimates chemical reactions between pollutants, and an inventory of pollutants emissions to compute concentration estimates.
The sources of error in this model are specifically analyzed by \cite{SIRANE-3} \textcolor{black}{in a study that compares SIRANE estimations against hourly measurements of 15 reference stations in Lyon (France) during 2008. This article lists } high \no2 concentrations and atmospheric conditions instability \textcolor{black}{as potential sources of errors}. Examples of SIRANE outputs are provided in Appendices \ref{app:outputs} \textcolor{black}{and \ref{app:correctedmaps}}.

\subsection{Air quality measurements}
\label{ssec:data-measurements}

\begin{figure}[htb]
    \centering
    \includegraphics[width=0.7\linewidth]{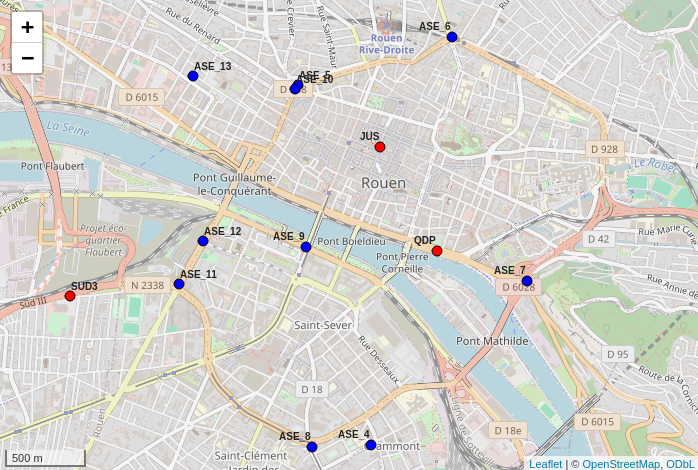}
    \caption{Map of Rouen provided by OpenStreetMap and locations of low-cost sensors (blue circles) and reference stations (red circles).}
    \label{fig:carte-ase}
\end{figure}

The air quality data used in this study is a subset of an openly available dataset fully described in \cite{DIB_Rouen2024}. It consists of hourly measurements from 10 low-cost sensors and 4 reference stations in Rouen, France, during December 2022. This period was chosen because \no2 concentrations tend to be higher in winter (see \cite{JuradoAssessment2020} for a recent study in France). The reference stations (model: HORIBA APNA 370) measure \no2 concentrations in \unit\ using chemiluminescence. Two stations (SUD3 and QDP) are installed at traffic sites, and the others (CHS and JUS) at background sites. The low-cost sensors installed in Rouen are AirSensEUR sensors (\cite{ASE_TechA, ASE_TechB}). They use metal oxide sensors that trigger electrochemical reactions with pollutants, generating an electric current proportional to their concentrations. They thus provide \no2 measurements in $\mu$V, but also nitrogen monoxide (NO in $\mu$V), ozone (O$_3$ in $\mu$V) and carbon monoxide (CO in $\mu$V). In addition, they are equipped with atmospheric pressure (in hPa), temperature (in \celsius C) and relative humidity (in \%) sensors. Each sensor is installed on a traffic light. The locations of the reference stations and low-cost sensors in the city are given in Figure \ref{fig:carte-ase}. One of the station (CHS) is excluded from this close-up map because it is far from the city center. Prior to deployment at their current sites, each low-cost sensor was collocated with a traffic monitoring station during winter 2021. \textcolor{black}{Thus, the sensors were one year old at the time of the study. As suggested by \cite{LiCharacterizing2021}, their behavior may differ from that during the collocation period. However, this period is neither useful nor necessary for the proposed approach, and only serves to provide the comparative values given in the appendix.}

\begin{figure}[H]
    \centering
    \includegraphics[width=\linewidth,trim = {0 0 0 0}, clip]{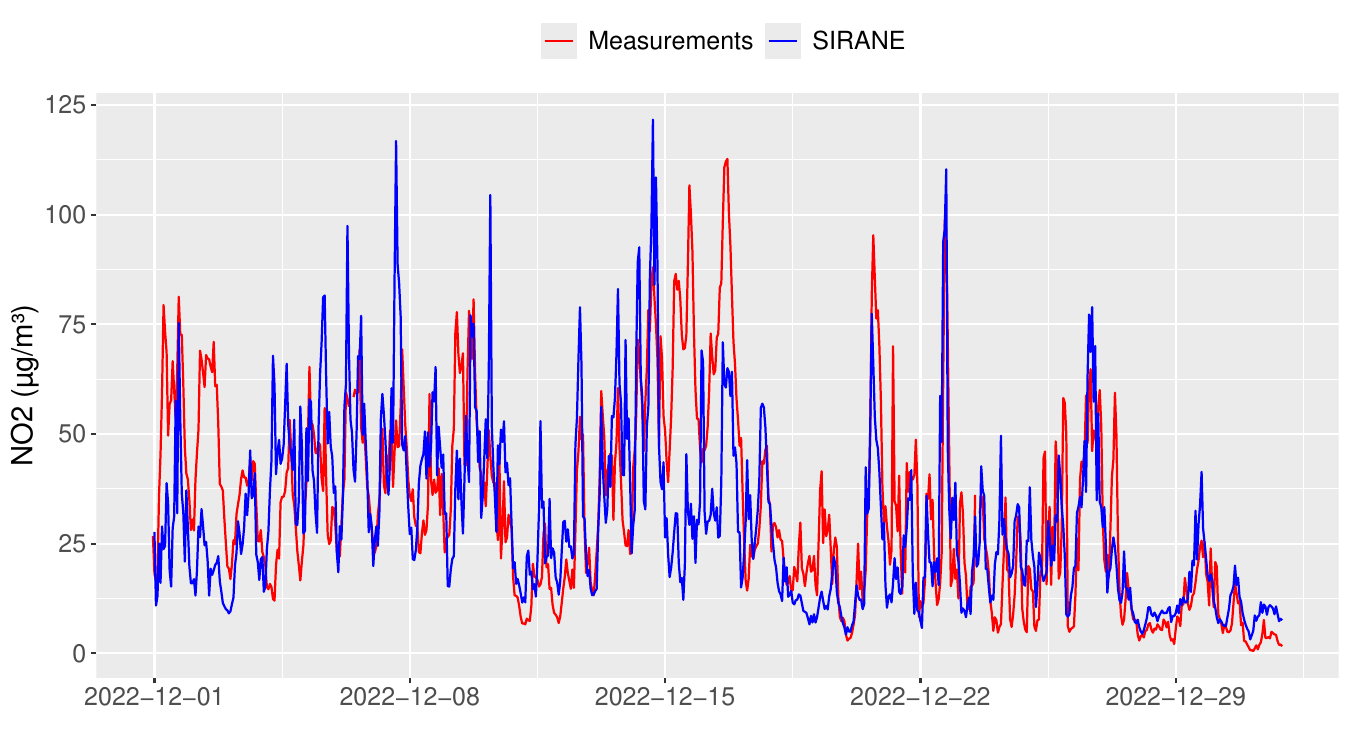}
    \caption{\no2 concentration at stations SUD3. Measurements are represented in red and SIRANE estimations in blue.}
    \label{fig:mesuresSUD3}
\end{figure}

\begin{figure}[H]
    \centering
    \includegraphics[width=\linewidth]{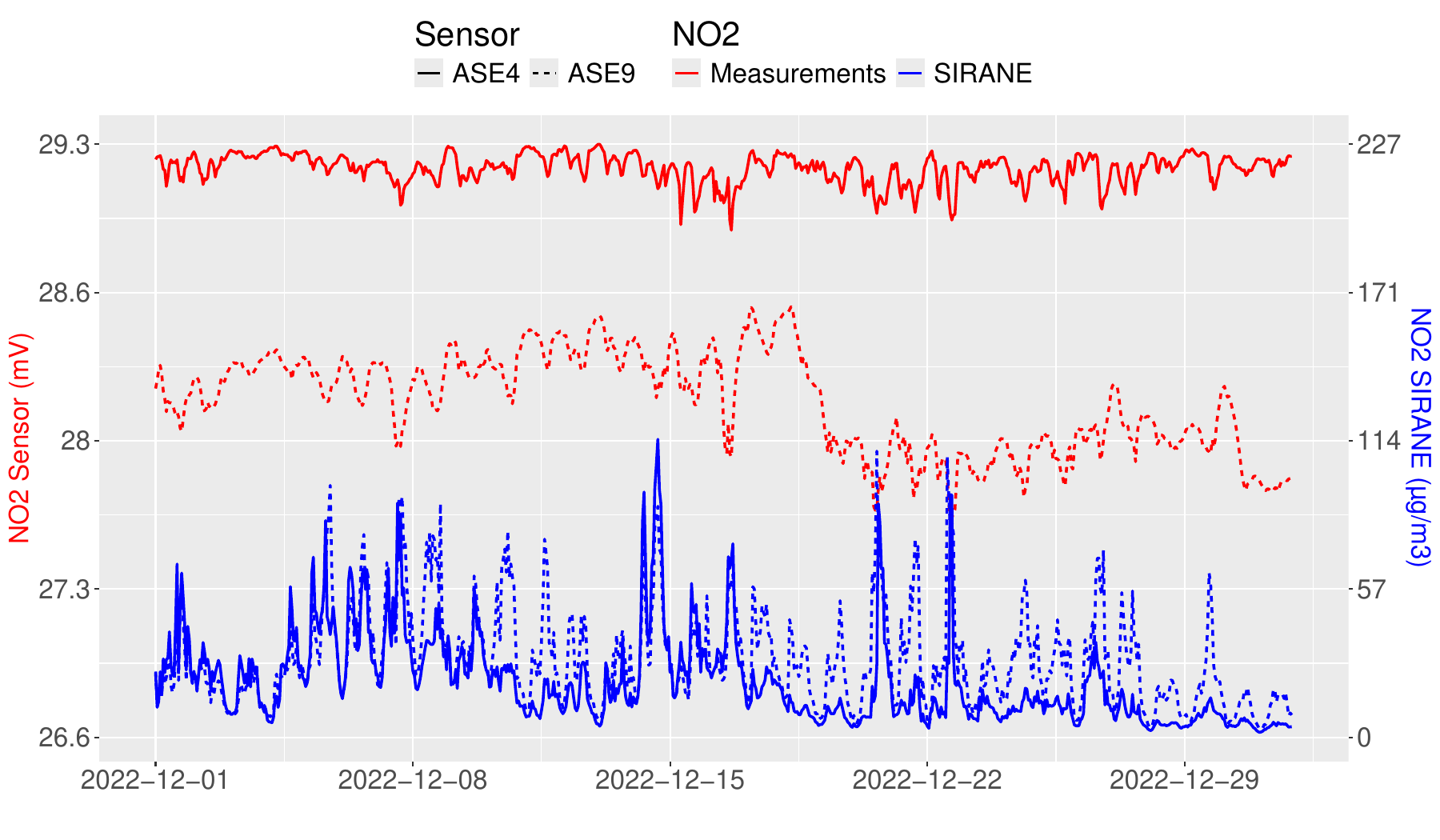}
    \caption{\no2 concentrations at sensors ASE4 \textcolor{black}{(solid lines)} and ASE9 \textcolor{black}{(dashed lines)}. Measurements (in mV) are represented in red (scale on left) and SIRANE estimations in blue (scale on the right).}
    \label{fig:mesuresASE4-9}
\end{figure}

Figure \ref{fig:mesuresSUD3} displays time series of measurements and SIRANE estimates at station site SUD3. Both series exhibit similar patterns, particularly when \no2 concentrations are low, but \textcolor{black}{sometimes} show significant discrepancies during concentration peaks, at which SIRANE \textcolor{black}{may} provides weaker concentrations \textcolor{black}{(see for example on the 2nd, 15th and 16th of December)}. Figure \ref{fig:mesuresASE4-9} presents time series from low-cost sensors ASE4 and ASE9 together with their corresponding SIRANE estimates (i.e the time series of the concentrations given by the SIRANE maps at the same location). The left axis corresponds to sensor measurements (mV), while the right axis corresponds to SIRANE estimates (\unit). \textcolor{black}{Both solid line curves (as well as dotted line ones)} can therefore be compared, as they both provide information related to the \no2 concentration at the same location. The figure illustrates that the electrical tension is a decreasing function of the actual \no2 concentration : lower voltages correspond to higher pollutant levels. It also highlights key challenges associated with low-cost sensors. First, the voltage scale may differ from one sensor to another (approximately 29 mV for ASE4 and 28 mV for ASE9), underscoring the need to build a separate calibration model for each device. Second, sensor behavior can change over time. For example, ASE9 exhibits a mid-period decrease of about $0.3$ mV. Consequently, a model calibrated over a specific period may no longer be valid at a later date, which emphasizes the need for in-situ calibration procedures.

\subsection{Spatial covariates}

In addition to the previously described data, the modeling framework incorporates covariates represented by spatial maps. The purpose of these variables is to account for land-use characteristics in the bias of the physico-chemical simulations, as suggested by \cite{LUR-Taiwan-2016} and \cite{trees-madrid}. Specifically, we extract from OpenStreetMap\footnote{see \cite{OpenStreetMap}} the surface of green areas within 50m of each point of the study area, and the surface of major roads within the same radius, both expressed in hm$^2$. \textcolor{black}{These information could be retrieved using other maps, such as CORINE Land Cover\footnote{see \cite{CORINE2018}}. However, OpenStreetMap provides multiple other information that were not included in this study, and operates through open collaboration; it is therefore often updated to reflect the current state of any city.}

\begin{figure}[ht]
    \centering
    \begin{subfigure}{0.49\linewidth}
        \centering
        \includegraphics[width=\textwidth, trim={1.3cm 0.6cm 0 0},clip]{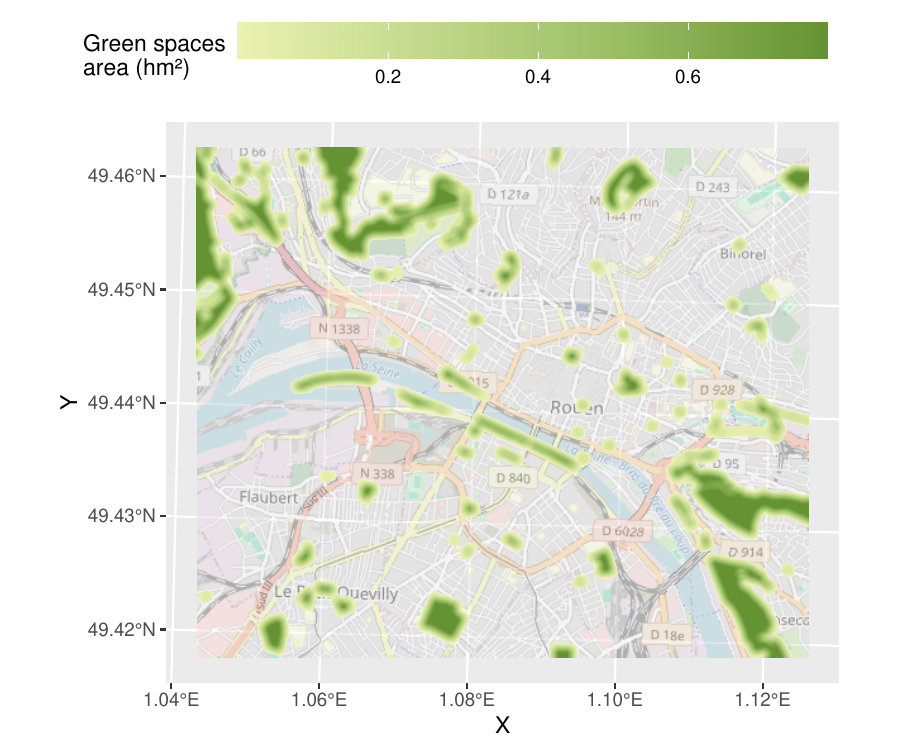}
        \caption{Green spaces area (hm$^2$)}
    \end{subfigure}
    \hfill
    \begin{subfigure}{0.49\linewidth}
        \centering
        \includegraphics[width=\textwidth, trim={1.3cm 0.6cm 0 0},clip]{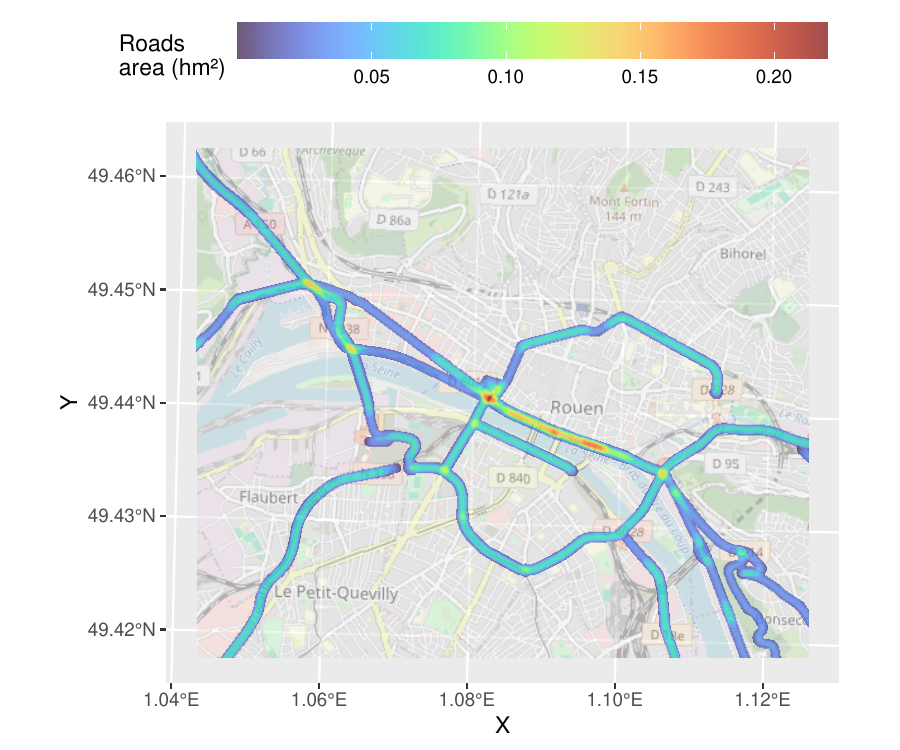}
        \caption{Roads area (hm$^2$)}
    \end{subfigure}
    \\
    \begin{subfigure}{0.49\linewidth}
        \centering
        \includegraphics[width=\textwidth, trim={1.3cm 0.6cm 0 0},clip]{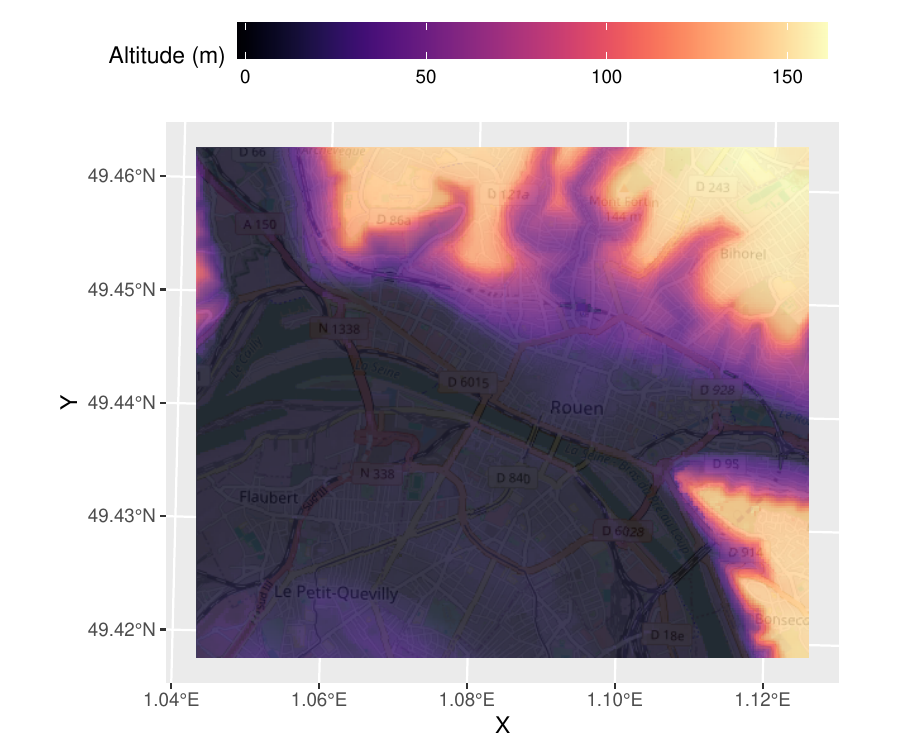}
        \caption{Elevation (m)}
    \end{subfigure}
    \caption{Spatial data in Rouen}
    \label{fig:cartes-GR}
\end{figure}

We also include an elevation map with a $25 \times 25$ m resolution, \textcolor{black}{part of the BD ALTI\footnote{see \cite{IGN}} database}, which we interpolate to match the $10 \times 10$ m resolution of the other spatial layers. Figure \ref{fig:cartes-GR} presents the maps of green space density, road density, and elevation across the study area. The choice of these covariates to explain errors in physico-chemical models outputs results from both a literature review and a preliminary analysis of the data. This part of the study must be adapted to the physico-chemical model and the city in question.

\subsection{Temporal covariates (Meteorological data)}
\begin{figure}[H]
    \begin{subfigure}{0.5\linewidth}
        \centering
        \includegraphics[width=\textwidth]{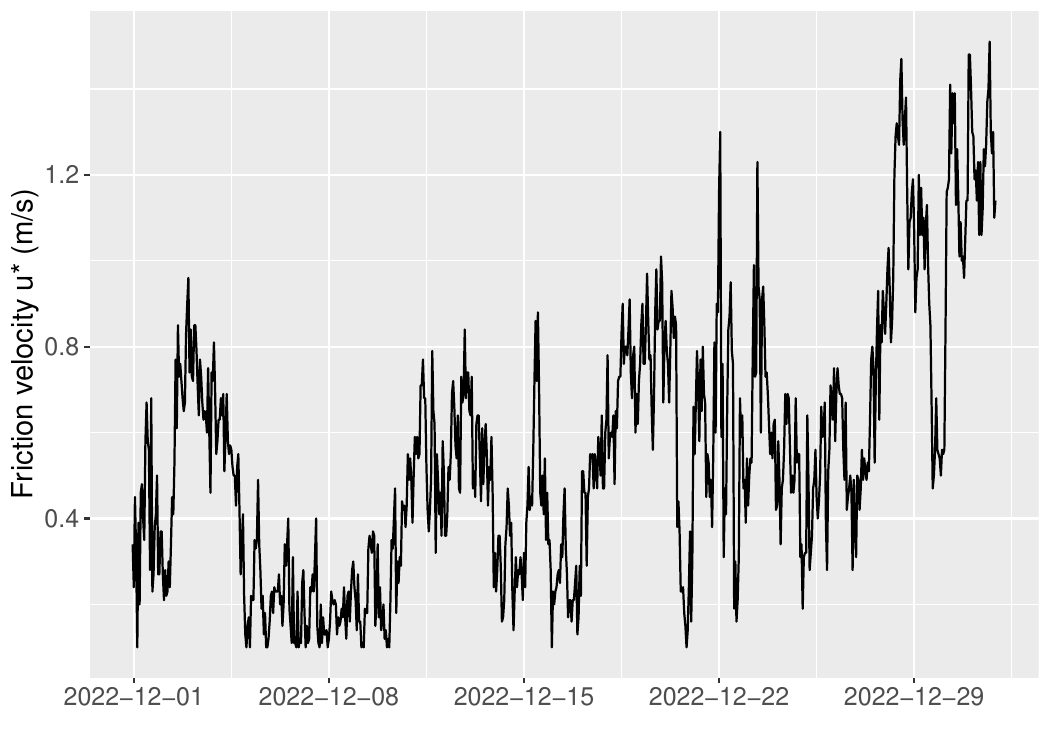}
        \caption{Friction velocity $u^*$ (in $m\cdot s^{-1}$)}
    \end{subfigure}
    \hfill
    \begin{subfigure}{0.5\linewidth}
        \centering
        \includegraphics[width=\textwidth]{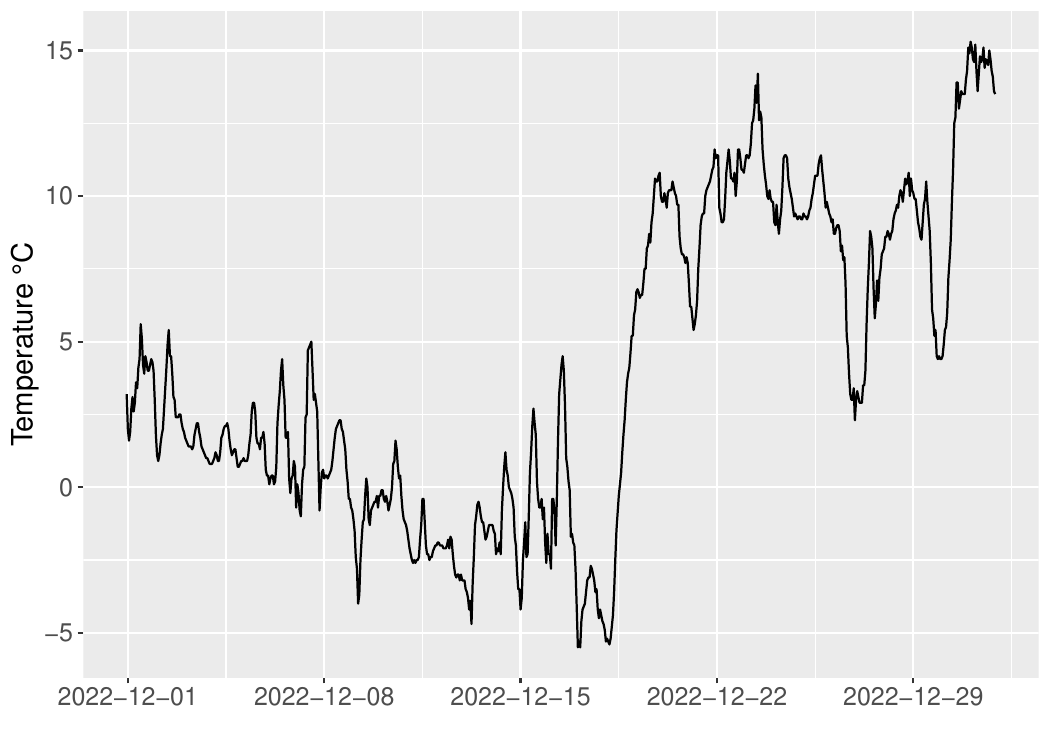}
        \caption{Temperature (in \celsius C)}
    \end{subfigure}
    \caption{Temporal data in Rouen during December 2022\textcolor{black}{, part of SIRANE's inputs. Retrieved from a local meteorogical monitoring station.}}
    \label{fig:var-temp}
\end{figure}

Meteorological variables, such as temperature, are already included as inputs in the computation of SIRANE outputs. However, due to city-specific geographic features, they may still contribute to errors of this physico-chemical model (\cite{SIRANE-2}). \textcolor{black}{Moreover, in the version of SIRANE that we have used, meteorological data were included as 1D variables (one value for the entire city at each moment).} In this study, we consider two temporal covariates: temperature (\celsius C) and \textcolor{black}{teh inverse of the} friction velocity $u^*$ ($\text{m}\cdot\text{s}^{-1}$). The latter is a standard quantity in fluid mechanics (see \cite{ustar}), that is used to represent shear-related motion in fluids. This quantity is known to be linked to wind speed, which plays a major role in the dynamics of pollutant concentrations. Both variables are measured at a meteorological station in Boos, located $10$ km from Rouen. They are not treated as spatially varying variables, since the study domain is sufficiently small to assume spatial constancy. An attempt to interpolate temperature from low-cost sensors and reference stations using ordinary kriging confirmed the absence of spatial dependence in the variogram. Figure \ref{fig:var-temp} shows the time series of both temperature and friction velocity during December 2022 in Rouen.

\section{Model}\label{sec:model}

\textcolor{black}{Our approach assumes that outputs from physico-chemical air quality models may exhibit systematic bias whose form is city-specific, because some geographical and meteorological characteristics of the urban environment (e.g. elevation, land use, population density) are only partially represented in these models and can besides interact.
The main novelty of our approach lies in the joint use of reference monitoring stations and low-cost sensors to infer a city-specific statistical correction of these physico-chemical model outputs. This allows us to capture local interactions between geographical, social and meteorological variables that are not explicitly represented in the physico-chemical model.
Our method provides a flexible way to adapt model outputs to the specific characteristics of a given city while preserving the general structure of the physico-chemical modelling framework.} \textcolor{black}{In \cite{AuderDebiaising2024} we studied a similar framework based on spatial partitioning, we generalize this idea by incorporating land-use variables, thereby overcoming the limitations of discrete partitioning and achieving spatially smooth corrections.}

\subsection{Bias of the physico-chemical model output}
\label{subsec:bias}

Several models are available for urban air quality assessment, including ADMS-Urban (\cite{ADMS-Urban}), MUNICH (\cite{KimMUNICH2018}) or SIRANE (\cite{SIRANE-1,SIRANE-2,SIRANE-3}). SIRANE estimates \no2 concentrations at an hourly scale a spatial resolution of $10\times 10$m by accounting for land use (e.g. road network, building shapes, street canyons, traffic density), chemical reactions between pollutants, background concentrations, and fluid mechanics principles. However, it remains subject to bias, particularly at high \no2 levels.

Let $\mathscr{D}\subset\mathbb{R}^2$ denote the spatial domain and $\mathscr{T}\subset\mathbb{R}^+$ the temporal domain, with $s\in \mathscr{D}$ a location and $t\in\mathscr{T}$ a time. Since the physico-chemical model provides an estimate of pollutant concentrations, its bias at $(s,t)$ is defined as

\begin{equation}
    M(s,t) = C(s,t) + B(s,t)
    \label{eq:biais_model}
\end{equation}
where $M$ is the model output, $C$ the true concentration and $B$ the bias to be estimated. 

The bias $B(s,t)$ is assumed to depend on covariates of three types: spatial (hereafter denoted $X_S(s)\in\mathbb{R}^k$, $s\in\mathscr{D}$), temporal (denoted $X_T(t)\in\mathbb{R}^l$, $t\in\mathscr{T}$) and spatio-temporal (e.g., pollutants concentrations, which depend jointly on $s$ and $t$). In this application, we consider $k=3$ spatial covariates: green spaces (known from \cite{trees-madrid} to reduce \no2\ concentrations, but not explicitly included in SIRANE's inputs); road density (as traffic is the main source of \no2) and elevation (particularly relevant in Rouen, located in a high-altitude basin). We also retain $l = 2$ temporal covariates: temperature and friction velocity, selected from SIRANE inputs as explanatory variables for bias at station locations. 

Our model for the bias of the physico-chemical outputs is given in Equation \eqref{eq:biais}, where, $\zeta_S \in \mathbb{R}^{k}$ is the vector of coefficients associated with the $k$ spatial covariates, and $\zeta_T, \theta_T \in \mathbb{R}^{l}$ are two vectors of coefficients associated with the $l$ temporal variables. Finally, $a_0$ and $a_c$ are scalar parameters.
\begin{equation}
    B(s,t) = a_0 + \theta_T\, X_T(t) + C(s,t) \left(a_c + \zeta_S \, X_S(s) + \zeta_T\, X_T(t)\right).
    \label{eq:biais}
\end{equation}
For readability purposes, the bias might as well be written
\begin{equation*}
    B(s,t) = L_0(s,t) + C(s,t)L_c(s,t),
\end{equation*}
with 
$$\begin{cases}
    L_0(s,t) &:= a_0 + \theta_T\, X_T(t)\\
    L_c(s,t) &:= a_c + \zeta_S \, X_S(s) + \zeta_T\, X_T(t)
\end{cases}.$$
This model was chosen in order to ensure that spatial covariates \textcolor{black}{play a specifically important role} when \no2\ concentrations are high. Several models linking concentration, covariates and physico-chemical model outputs have been tested and this one gave the best results for the city of Rouen. \textcolor{black}{Similarly, the choice of the explanatory covariates must depend on the city of interest. For the specific case of Rouen, the selection of covariates resulted from a preliminary study of the data and the bibliography.}

\subsection{Measures}

Recall from Section \ref{sec:data} that 2 types of data are available: measurements from $I$ reference stations and $J$ low-cost sensors. For any $i\in\{1,...,I\}$ and any $j\in\{1,...,J\}$, let $s_{0,i}\in\mathscr{D}$ denote the location of reference station $i$ and $s_{1,j}\in\mathscr{D}$ the location of low-cost sensor $j$. While locations could, in principle, vary over time, this is not the case for the dataset analyzed here. Measurements from reference stations are considered reliable, whereas low-cost sensors may produce erroneous values due to various factors and typically have a limited lifespan of about two years. These errors might arise from environmental conditions, meteorological factors, interference from other pollutants, or sensor aging (\cite{WeiImpact2018, LiCharacterizing2021,ChristakisAir2023}). Moreover, low-cost sensors provide raw signals (in $\mu V$), that must be converted into pollutant concentrations. As noted in Section \ref{sec:data}, they also record additional variables such as temperature and relative humidity, which can be incorporated to explain their \no2 concentration measurements. \cite{BobbiaStatistical2025} shows that the relationship between sensor outputs and true concentrations is device-specific and is well approximated by a multiple linear model. Let $Z_{0,i}(t)$ (resp. $Z_{1,j}(t)$) denote the measurement made by the reference station $i$ (resp. the low-cost sensor $j$) at time $t$. Accordingly, we adopt the following model where $j$ denotes the sensor index and $Y_j(t)$ denotes the covariates measured by the sensor.

\begin{align}
    Z_{0,i}(t) &= C(s_{0,i},t) + \varepsilon_0(i,t), & \text{with }\varepsilon_0(i,t)\sim \mathcal{N}(0,\sigma^2_0)
    \label{eq:mes_sta}\\
    Z_{1,j}(t) &= f\left(C(s_{1,j},t),j,Y_j(t)\right)+\varepsilon_1(j,t), & \text{with }\varepsilon_1(j,t)\sim \mathcal{N}(0,\sigma^2_j)
    \label{eq:mes_mc}
\end{align}
where \begin{equation}\label{eq:mc}f(C,j,Y)=\beta_j+\alpha_j C+\gamma_j Y\end{equation} and the parameters $\beta_j\in\mathbb{R}$, $\gamma_j\in\mathbb{R}^q$, $\alpha_j\in\mathbb{R^*}$ are to be estimated. We moreover assume that the random variables $((\epsilon_0(i,t)$, $\epsilon_1(j,t)), i\in\{1,...,I\}, j\in\{1,...,J\}, t\in\mathscr{T})$ are all independent.

Equations \eqref{eq:mes_mc} and \eqref{eq:mc} imply that the $\alpha_j$ ($j\in\{1\ldots J\}$) parameters shall be negative, since as mentioned in section \ref{ssec:data-measurements}, the electric tension is negatively correlated with \no2 concentrations. Additional prior information can be derived from previous sensor studies. For example, an acceptable \textcolor{black}{order of magnitude} for each \textcolor{black}{calibration} parameter can be obtained by studying the calibration model of a sensor during a collocation period, when sensors are installed at stations sites. A more thorough discussion concerning prior information on parameters will be developed in the Subsection \ref{sec:priors}.

Finally, it is possible to introduce a time dependence to parameters $\alpha_j$, $\beta_j$, $\gamma_j$ and $\sigma_j$, as suggested by the orange curve of Figure \ref{fig:mesuresASE4-9}. This would require a longer period of measurements to confirm the necessity of time-varying coefficients.

\section{Estimation}\label{sec:estim}

\subsection{Parameters}

Equations \eqref{eq:biais_model} to \eqref{eq:mc} define the proposed probabilistic model, linking physico-chemical model outputs, pollutant concentrations, spatial and temporal covariates, and measurements from both reference stations and low-cost sensors. The objective is to estimate two sets of parameters: $a_0, a_c\in\mathbb{R}$, $\theta_T, \zeta_T\in\mathbb{R}^l$, $\zeta_S\in\mathbb{R}^k$ which characterize the bias of the physico-chemical model (Equation \eqref{eq:biais_model}), and $\alpha_j\in\mathbb{R}^*$, $\beta_j\in\mathbb{R}$ and $\gamma_j\in\mathbb{R}^q$, which describe the behavior of each low-cost sensor. In total, we aim to estimate $p = 3 + k + 2l + J\times (q + 2)$ parameters. From Equations \eqref{eq:biais_model} and \eqref{eq:biais}, we obtain, for any $s\in\mathscr{D}$ and $t\in\mathscr{T}$,
$$M(s,t)=a_0+\theta_TX_T(t)+C(s,t)(1+a_c+\zeta_SX_S(s)+\zeta_TX_T(t)),$$ therefore from Equation \eqref{eq:mes_sta}, for any $i\in\{1,...,I\}$,
\begin{equation}
\begin{split}M(s_{0,i},t)&=a_0+\theta_TX_T(t)+(Z_{0,i}(t)-\epsilon_0(i,t))(1+a_c+\zeta_SX_S(s_{0,i})+\zeta_TX_T(t)),\\&=a_0+\theta_TX_T(t)+Z_{0,i}(t)(1+a_c+\zeta_SX_S(s_{0,i})+\zeta_TX_T(t))+\epsilon'_0(i,t)\end{split}
\label{eq:modele-stations}
\end{equation}
where $$\varepsilon_0'(i,t) \sim \mathcal{N}\left(0,\big(1 + a_c+\zeta_S X_S(s_{0,i})+\zeta_TX_T(t)\big)^2 \sigma_0^2\right),$$ which links linearly the measurements taken by reference stations and the physico-chemical outputs at reference station locations.
Similarly, from Equations \eqref{eq:mes_mc} and \eqref{eq:mc}, for any $j\in\{1,\ldots,J\}$,%
\begin{equation}
\begin{split}
M(s_{1,j},t)&=a_0+\theta_TX_T(t)\\
&+\frac{1}{\alpha_j}\left(Z_{1,j}(t) - \beta_j - \gamma_j Y_j(t)-\varepsilon_j(t)\right)(1+a_c+\zeta_SX_S(s_{1,j})+\zeta_TX_T(t))\\
&=a_0+\theta_TX_T(t)\\&+\frac{1}{\alpha_j}\left(Z_{1,j}(t) - \beta_j - \gamma_j Y_j(t)\right)(1+a_c+\zeta_SX_S(s_{1,j})+\zeta_TX_T(t))\\
&+\varepsilon'_1(j,t),
\label{eq:modele-capteurs}
\end{split}
\end{equation}
where $$\varepsilon_1'(j,t) \sim \mathcal{N}\left(0,\big(1 + a_c+\zeta_S X_S(s_{1,j})+\zeta_TX_T(t)\big)^2 \sigma_j^2\right).$$
Since Equation \eqref{eq:modele-stations} is a particular case of Equation \eqref{eq:modele-capteurs}, the link between the physicochemical model output and measures can be summarized by the single Equation \eqref{eq:modele-capteurs}, but assuming that for reference stations, $\alpha_0 = 1$, $\beta_0 = 0$ and $\gamma_0=\{0\}^q$.

\subsection{Estimation procedure}

From Equation \eqref{eq:modele-stations}, reference station data alone enable the estimation of the parameters $\theta_T$, $a_0$, $a_c$, $\zeta_S$, $\zeta_T$ and $\sigma_0$. However, the heteroskedasticity induced by the non-constant variance of $\varepsilon'_0(i,t)$ requires the use of generalized least squares (GLS) estimators (\cite{goldberger1972fgls}). Once these parameters are obtained, low-cost sensor data, as described in Equation \eqref{eq:modele-capteurs}, can be used to estimate the parameters $\alpha_j$, $\gamma_j\in\mathbb{R}^q$, $\beta_j$ and $\sigma_j$ for all $j$. Nevertheless, this approach does not exploit the low-cost sensor measurements to improve bias estimation. 

Our goal is to combine data coming from both reference stations and low-cost sensors to simultaneously estimate the calibration parameters of each low-cost sensor and the parameters governing the bias of the physico-chemical model. \textcolor{black}{More specifically, reference stations alone allow for a statistical correction of the bias of the physico-chemical model. However we expect that due to the large number of low-cost sensors, combining both datasets simultaneously allows for obtaining a better correction of this bias, and for estimating calibration parameters of low-cost sensors, i.e. for understanding their behavior and the evolution of this behavior with time.} Since the likelihood of the model parameters given all the data cannot be obtained analytically, we adopt a Bayesian inference framework. Conditional on fixed parameters, the nested structure of the model allows for efficient data simulation, enabling posterior inference through Markov Chain Monte Carlo (MCMC) sampling (see \cite{metropolis1953equation}). The procedure is implemented using JAGS\footnote{JAGS stands for 'Just Another Gibbs Sampler'}, which was first described by \cite{Plummer2003JAGS}.

Bayesian inference treats model parameters as realizations of random variables (see \cite{Lynch2007Bayesian} or \cite{Lee2012Bayesian}). This framework is particularly useful when prior information about the parameters is available, as in the present context. For example, since all sensors are of the same brand, they are \textcolor{black}{\textit{a priori} exchangeable}, which justifies using \textcolor{black}{the same} prior distribution for parameters $\alpha_j,\beta_j, \gamma_j$, and $\sigma_j$ across sensors. Moreover, prior knowledge about the sign or range of \textcolor{black}{some} parameters -- for instance, the calibration coefficients $\alpha_j$, expected to be negative -- can be explicitly incorporated. When calibration has already been performed on some sensors, the resulting coefficients provide highly informative priors for subsequent analysis.

Using both data and prior information, the posterior distribution of the parameters is approximated via MCMC algorithms. In this study, Gibbs sampling is employed to generate posterior draws. Let $\Theta$ denote the vector of parameters to be estimated (bias parameters, sensors parameters and standard deviations). At iteration $m+1$, the $k$-th parameter is sampled from its full conditional distribution:
\begin{equation}
    \Theta_k^{(m+1)} \sim \pi\left(\Theta_k | \Theta_1^{(m+1)}, \dots , \Theta_{k-1}^{(m+1)}, \Theta_{k+1}^{(m)}, \ldots, \Theta_p^{(m)}\right)\quad\text{when } k\in\{1\ldots,p\}
\end{equation}
where $\pi$ is obtained from Bayes' theorem and the prior distributions. Initial values $\Theta^{(0)}_k$, $k\in\{1,\ldots,p\}$ can be sampled from the priors or specified directly; in this application, sensor parameters are initialized from coefficients estimated during a previous collocation period (see Table \ref{tab:coloc_coefs} in Appendix). Convergence properties of Gibbs sampling are established (see for example \cite{Gaetan_Guyan_Gibbs2008}). The algorithm is widely implemented, including in JAGS, which we use via the \texttt{rjags} package in \texttt{R}. MCMC and Gibbs sampling are particularly well suited here as they naturally accommodate hierarchical modeling.

In the present setting, the Bayesian model is defined as follows. At iteration $m$, parameter values are drawn from their prior distributions. Then, for each time $t$ and device location $s_k$:

\begin{align} 
    & \begin{cases}
        L_0^{(m)}(s_k,t) &= a_0^{(m)} + \theta_T^{(m)} X_T(t)\\
        L_c^{(m)}(s_k,t) &= a_c^{(m)} + \zeta_S^{(m)} X_S(s_k) + \zeta_T^{(m)} X_T(t)\\    
    \end{cases}\\
   & C^{(m)}(s_k,t) = \frac{M(s_k,t) - L_0^{(m)}(s_k,t)}{1 + L_c^{(m)}(s_k,t)}& \label{eq:sim-biais}\\
   & Z_{k}^{(m)}(t) = \alpha_k^{(m)}C^{(m)}(s_k,t) + \beta_k^{(m)} + \gamma_k^{(m)} Y_k(t) + \varepsilon^{(m)}(k,t) & \label{eq:sim-mesures}
\end{align}

The simulated measurements $Z_{k}^{(m)}(t)$ can then be compared with the observed values $Z_{k}(t)$. 

The Gibbs sampler yields posterior distributions for all model parameters. The Bayesian estimator $\hat{\theta}$ of a parameter $\theta$ can be defined either as $\hat{\theta}^\text{MSE}$, the posterior mean of the samples $\{\theta^{(1)} ... \theta^{(n)}\}$, which minimizes the mean square error, or as $\hat{\theta}^\text{MAP}$, the posterior mode. In this study, we retain $\hat{\theta}^\text{MSE}$ but since the posterior distributions are unimodal and concentrated, both estimators provide similar results.

All computations were performed with 8000 adaptation iterations before sampling, 2000 samples discarded for the burn-in phase and 2500 samples drawn from the posterior distribution and retained. Convergence was assessed by computing multiple chains in parallel.

\begin{figure}[htb]
    \centering
    \includegraphics[width=\linewidth, trim={0cm 5cm 0 5cm},clip]{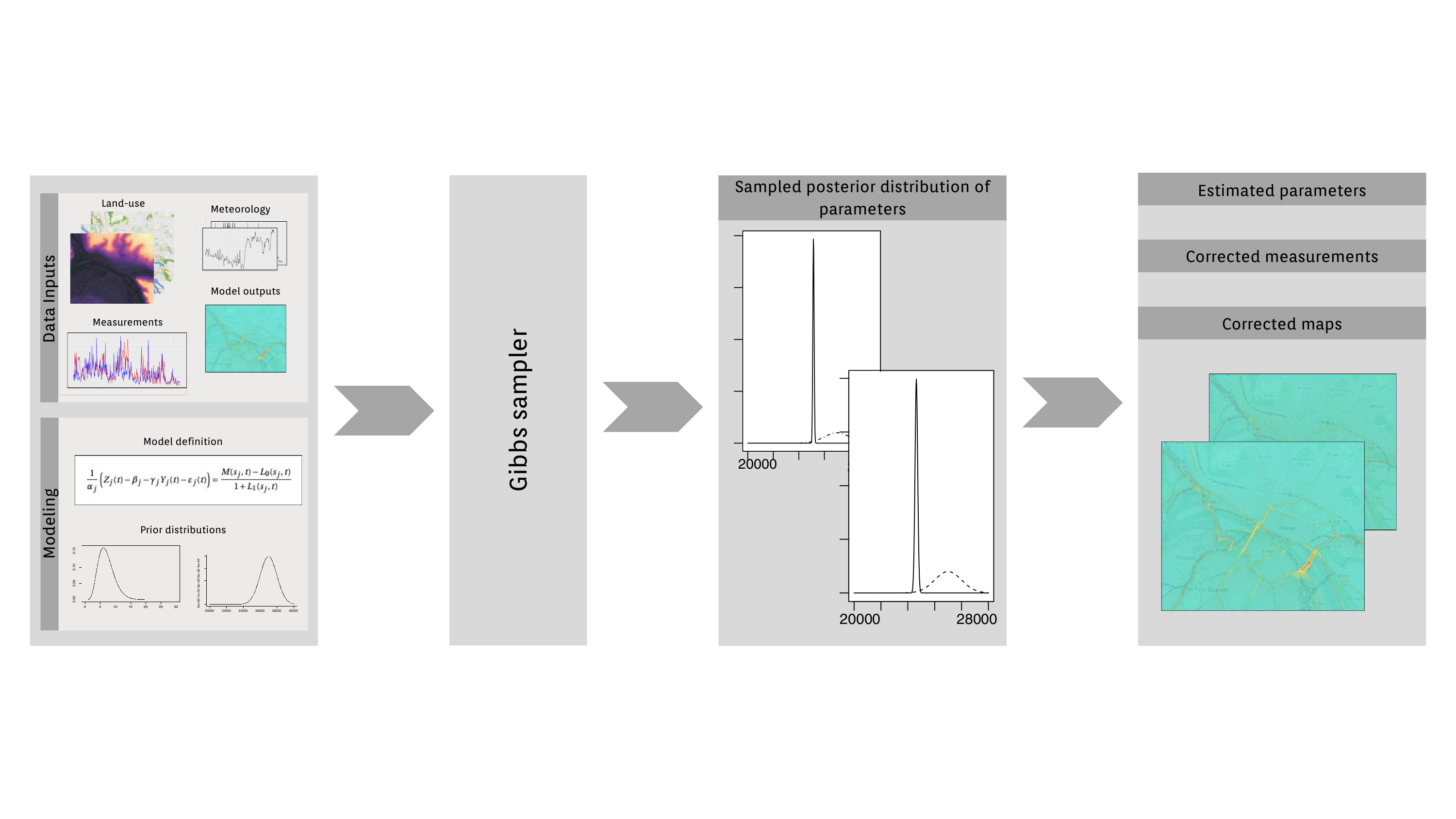}
    \caption{\color{black} Scheme of the estimation process}
    \label{fig:scheme}
\end{figure}

\textcolor{black}{A scheme of the entire estimation process is given in Figure \ref{fig:scheme}.}

\subsection{Prior distributions}\label{sec:priors}
This section details the prior information used to estimate bias and calibration parameters. The choice of prior distributions is guided by observations of the sensors behavior and physical reasoning.
For sensors, the parameters $\alpha_j$ are expected to be negative, reflecting the negative correlation between raw measurements (in mV) and concentrations (in \unit). Standard deviations $\sigma_j$ are strictly positive and are assigned a Weibull prior distribution, enabling a wide range of values. Regarding the bias parameters, the parameter $a_c$ is expected to be negative as SIRANE tends to underestimate high \no2 concentrations (\cite{SIRANE-3}). The same argument implies that among the spatial coefficients, $\zeta_{S,1}$ (coefficient associated to roads) should be negative, as road proximity tends to increase \no2 concentrations. In contrast, $\zeta_{S,2}$ (coefficient associated to green spaces) should be positive, as concentrations are generally lower in these areas, and their presence is not taken as input in SIRANE's model. Finally, $\zeta_{S,3}$ (coefficient associated with elevation) is also expected to be positive, reflecting the specific topography of Rouen, where pollution tends to be less concentrated at higher elevations. 

For parameters that need to be negative ($\alpha_j$ and some bias parameters), we propose that their opposites follow a gamma prior distribution\textcolor{black}{. This distribution is commonly used in Bayesian statistics to deal with positive random variables as it is highly versatile thanks to its two parameters}. Finally, other distributions are supposed normal, and rather uninformative. More information on prior distributions and the possible values for low-cost sensors parameters are given in Appendix \ref{app:priors}.

\subsection{Validation process}
\label{subsec:validation}
Our main objective is to improve \no2 concentration maps, i.e. to produce pollutant concentration estimations that are closer to real concentrations, or to measurements made by monitoring stations. We focus only on hours that are most likely polluted (traffic hours, between 6 a.m. and 10 a.m. and between 4 p.m. and 8 p.m. \textcolor{black}{and during weekdays (from Monday to Friday)}). That ensures that we do not have too many data points representing periods with no signal (\no2 concentrations being very low during night\text{ and weekend}). \textcolor{black}{We call $\mathscr{T}$ this restrained set of hours.}

To evaluate the performance of our method, \textcolor{black}{we proceed by two manners. First, we} divide $\mathscr{T}$ into two \textcolor{black}{sets}: 70\% of the instants are sampled and the associated data are used to train the model, and the other 30\% are used to test the reliability of the estimations. \textcolor{black}{This allows us to evaluate if the estimated parameters are efficient to correct maps of unseen times but on position that have been included in the estimation procedure.}
\textcolor{black}{Second, we proceed by cross-validation, using a leave-one-out procedure. More precisely, it means that} for each monitoring station $i$, the bias parameters are estimated without using measurements $Z(s_{0,i},\cdot)$, and using only the training dataset. The corrected map is then computed $\forall t\in\mathscr{T}$ at location $s_{0,i}$. \textcolor{black}{This allows us to evaluate the capacity of the modeling to improve the maps at positions where no measurements are available.}
Finally, \textcolor{black}{for both situation,} we assess \textcolor{black}{the} performances by studying the distance between the corrected concentrations $M'(s_{0,i},\cdot)$ and the measurements made by the remaining station $Z(s_{0,i},\cdot)$.

One of the main questions addressed in this study, which is a major challenge for air quality monitoring agencies, concerns the ability of low-cost sensors to contribute to improving concentration maps. Indeed, the proposed strategy can be achieved using only reference monitoring stations. While low-cost sensors provide a finer spatial representation, they also imply taking into account lower-quality data. The added-value of low-cost sensors is evaluated by comparing the concentration map correction performance with and without sensors measurements. This comparison is given in Table\textcolor{black}{s \ref{tab:performance-train-test} and} \ref{tab:perfo-cv}.

\section{Results}\label{sec:results}
This section focuses on the results obtained using our method. First, we present corrected concentration maps and interpret the estimated parameters, both for the bias and for the sensors. Then, we assess the quality of the method at locations where measurements are available \textcolor{black}{using the train and test sets} and where no measurement are available using a leave-one-out procedure. We finally compare the corrected map obtained by using only reference stations, with the one obtained by using both reference stations and low-cost sensors.

\subsection{Outputs of our approach}
Outputs of our approach include corrected maps, sensor calibration parameters, and estimates of the bias parameters. Figure \ref{fig:corrected-map} \textcolor{black}{in Appendix \ref{app:correctedmaps}} presents two instants of the dataset that are of interest to illustrate the correction of SIRANE outputs.

\textcolor{black}{Another aspect of interest is the respect of diurnal variation in the concentrations. To illustrate this question, we have computed the mean of measurements, SIRANE estimation and the corrected map estimations at each hour of the day and for each monitoring station. It has shown that the diurnal variation were preserved by the framework. An example, based on the reference station SUD3, is given in Figure \ref{fig:sud3_hour} of Appendix \ref{app:diurnalvar}.}

\textcolor{black}{Then,} let us focus on parameter estimates. Recall that there are 10 coefficients that need to be estimated for the bias. Table \ref{tab:coefficients} summarizes the estimated coefficients using only reference stations (row (S)) and both reference stations and low-cost sensors (row (S + LCS)). It can be noticed that in both cases, the estimated parameters are of the same order of magnitude, with some differences. Parameters $\sigma_0$ are almost equal in both cases. The estimation using low-cost sensors shows less importance for spatial parameters than the one using only reference stations, and greater importance for temporal parameters. Indeed, the coefficient $\theta_{T,1}$, associated with $1/u^*$, is greater (in absolute value) than the one obtained using only reference stations \textcolor{black}{(-7.75 v. -5.36)}. The same can be said for $\theta_{T,2}$, associated with temperature and $\zeta_{T,1}$. Regarding $\zeta_S$, it can be noted that the coefficient associated with elevation ($\zeta_{S,3}$) is the same in both estimations (0.01). As it is positive, the greater the elevation, the greater the bias. $\zeta_{S,2}$ is of \textcolor{black}{0.32} when estimated using low-cost sensors, which indicates that for a concentration of pollutant of 1 \unit, being located in a place surrounded by 1 hm$^2$ of green space accounts for \textcolor{black}{0.32} \unit of the bias. On the other hand, being surrounded by 1 hm$^2$ of roads accounts for an underestimation of the model by \textcolor{black}{-0.26} \unit.

\begin{table}[htb]
    \color{black}
    \centering
    \begin{tabular}{c|cccccc}
    & $a_0$ & $a_c$ & $\theta_T$ & $\zeta_S$ & $\zeta_T$ & $\sigma_0$ \\
    \hline
    (S) & -2.08 & -0.45 & $(-5.36,-0.25)^T$ & $(-0.35,0.40,0.01)^T$ & $(0.19,0.01)^T$ & 16.67 \\
    (S + LCS) & -1.88 & -0.67 & $(-7.75,-0.07)^T$ & $(-0.26,0.32,0.01)^T$ & $(0.31,0.02)^T$ & 16.76 \\
    \end{tabular}
    \caption{Estimated bias parameters, using reference stations and low-cost sensors.}
    \label{tab:coefficients}
\end{table}

Sensors correction coefficients are given in Table \ref{tab:coeffs-sensors}. A comparison can be made with coefficients obtained when the sensors were collocated in the winter of the previous year (see Table \ref{tab:coloc_coefs} in Appendix). It appears that the coefficients have changed, which supports the idea of re-calibrating sensors once deployed. Moreover, the coefficients are very different from one sensor to another, which suggests that each sensor should have its own calibration model. Finally, since the $\gamma$ coefficients are not null, it seems necessary to  account for other pollutants that can cause cross-sensitivity, as well as environmental factors (such as relative humidity or temperature in this case). Besides, \cite{LiCharacterizing2021} showed that \no2\ sensors measurements were less correlated to \no2\ concentration with time, which is supported by the estimated coefficients, which are lower (in absolute value) than they were one year before. On the other hand, coefficients associated with the measurement of Ox have increased from around 0.1 to around 0.3, which is consistent with the increased O$_3$ cross-sensitivity of the sensor as it ages.

\begin{table}[ht]
\color{black}
\centering
\begin{tabular}{r|ccccccc|c}
  \hline
Sensor & $\beta$ & $\alpha$ & $\gamma_\text{NO}$ & $\gamma_\text{CO}$ & $\gamma_\text{Ox}$ & $\gamma_\text{HR}$ & $\gamma_\text{T}$ & $\sigma$ \\ 
  \hline
ASE4 & 25728.71 & -1.58 & 0.00 & -0.06 & 0.21 & -1.68 & -4.33 & 33.07 \\ 
  ASE5 & 26998.27 & -1.60 & -0.07 & -0.04 & 0.24 & -1.35 & -2.04 & 31.90 \\ 
  ASE6 & 27083.11 & -1.36 & -0.07 & -0.03 & 0.20 & -0.03 & -24.48 & 35.49 \\ 
  ASE7 & 28042.68 & -1.07 & -0.07 & -0.01 & 0.16 & -0.95 & 2.97 & 34.74 \\ 
  ASE8 & 24003.27 & -2.22 & -0.10 & -0.01 & 0.33 & -1.43 & -8.50 & 33.41 \\ 
  ASE9 & 23561.47 & -2.49 & -0.10 & 0.00 & 0.32 & -1.48 & -29.70 & 34.20 \\ 
  ASE10 & 25990.97 & -1.37 & -0.07 & -0.02 & 0.25 & -1.39 & -10.64 & 33.36 \\ 
  ASE11 & 27646.75 & -1.21 & -0.08 & -0.02 & 0.19 & -1.34 & -23.69 & 32.63 \\ 
  ASE12 & 27241.07 & -1.39 & -0.07 & -0.02 & 0.20 & -1.93 & 2.31 & 31.56 \\ 
  ASE13 & 24870.06 & -1.78 & -0.04 & -0.04 & 0.24 & -0.87 & -33.86 & 31.08 \\ 
   \hline
\end{tabular}
\caption{Estimated coefficients of sensors calibration}
\label{tab:coeffs-sensors}
\end{table}

\subsection{Validation and low-cost sensors benefits}

\textcolor{black}{\subsubsection{Error evaluation}}

We assess the quality of correction using three indicators: the proportion of explained variance (EV), the mean absolute error (MAE), and the root mean squared error (RMSE). Considering $n$ measurements of \no2 concentrations at multiple reference stations and times, let $\tilde{M}(s,t)$ be the map of interest (either the initial map or corrected map) and $\bar{Z}_0$ be the mean of all considered $Z_{0,i}(t)$ values. The scores are defined in Equation \eqref{eq:scores}. Note that MAE and RMSE are better when they approach 0, while EV is better when it approaches 100.

\begin{equation}
\begin{split}
    \text{EV} &= 100\times\left(1 - \frac{\sum_{i,t}\left(Z_{0,i}(t) - \tilde{M}(s_{0,i},t)\right)^2}{\sum_{i,t}\left(Z_{0,i}(t) - \bar{Z}_0\right)^2}\right)\\
    \text{MAE} &= \frac{1}{n}\sum_{i,t}| Z_{0,i}(t) - \tilde{M}(s_{0,i},t)|\\
    \text{RMSE} &= \sqrt{\frac{1}{n}\sum_{i,t} \left(Z_{0,i}(t) - \tilde{M}(s_{0,i},t)\right)^2}\\
\end{split}
\label{eq:scores}
\end{equation}

\textcolor{black}{\subsubsection{Modeling evaluation on test and training sets}}

\color{black}
    First, the most obvious way to assess quality of a modeling is to compute the EV, MAE and RMSE both on the learning and the test sets. These values are provided in Table \ref{tab:performance-train-test} for three models: SIRANE, a correction using only reference stations (S) and a correction using both reference station and low-cost sensors (S + LCS). The objective is thus to understand how much the modeling can improve the initial physico-chemical model, but also to identify the added value of low-cost sensors in the estimation of the bias.

        \begin{table}[htb]
        \color{black}
        \centering
        \begin{tabular}{r|r|c|c|c}
            \hline
            & & EV & RMSE  & MAE \\
            \hline
            \multirow{3}{*}{Training set} & SIRANE & 7.5 \% & 20.5 & 15.4\\
            & Correction (S) & 38.2 \% & 16.7 & 12.9\\
            & Correction (S + LCS) & 37.7 \% & 16.8 & 13.0\\
            \hline
            \multirow{3}{*}{Test set} & SIRANE & 0.5 \% & 19.6 & 14.9 \\
            & Correction (S) & 40.9 \% & 15.1 & 11.7\\
            & Correction (S + LCS) & 41.1 \% & 15.1 & 11.7\\
            \hline
        \end{tabular}
        \caption{\color{black} SIRANE and correction performances (EV in \%, MAE in \unit\ and RMSE in \unit) evaluated on the training and test sets.}
        \label{tab:performance-train-test}
    \end{table}

    To analyze this table, it is important to keep in mind that the evaluation is made only where reference measurements are available, i.e. at reference stations. First on the training set, it appears that both correction models improve SIRANE's estimation, with a RMSE decreasing by almost 4 \unit. It seems that the use of low-cost sensors does not improve correction, which seems coherent given the fact that the evaluation set consists of training data from stations only. Thus, it appears more interesting to look at the evaluation using the test set. We can therefore notice that the correction performances are also similar in both cases, and similar to those on the training set. However, to evaluate the added-value of low-cost sensors, another situation must be considered, that is, the correction at an unobserved position. This situation is simulated by conducting a cross-validation in space.
\color{black}

\textcolor{black}{\subsubsection{Modeling evaluation by cross-validation}}
Recall that our objective is to improve SIRANE's estimations at locations $s$ where no measurement is available. As explained in the previous section, one way to represent this situation is by considering a leave-one-out procedure. For each station, SIRANE's bias is estimated without data from that station. Corrected estimations are then constructed using data from the most likely polluted hours. Performances are computed on all leave-one-out estimates and presented in Table \ref{tab:perfo-cv}.

\begin{table}[ht]
    \centering
    \color{black}
    \begin{tabular}{r|c|c|c}
    \hline
    & EV & RMSE & MAE\\ 
    \hline
    SIRANE & 13.4 \% & 18.6 & 13.7  \\ 
    Correction (S) & 25.3 \% & 17.2 & 13.2  \\ 
    Correction (S + LCS) & 33.1 \% & 16.3 & 12.6 \\ 
    \hline
    \end{tabular}
    \caption{Model output and corrections performances, computed with the leave-one-out procedure. RMSE and MAE are expressed in \unit.}
    \label{tab:perfo-cv}
\end{table}

In the leave-one-out situation, using both station and low-cost sensor measurements, the performance is satisfactory, with SIRANE's RMSE decreasing by \textcolor{black}{2.3} \unit\ \textcolor{black}{(12.4\%)} using our procedure. The percentage of explained variance increases from \textcolor{black}{13.4\% to 33.1\%}. It is thus reasonable to conclude that the proposed strategy helps correcting SIRANE's output where not measurements are available from monitoring stations.
However, when the bias is estimated using only monitoring stations, the estimation procedure \textcolor{black}{shows worse performances}. This can be explained by noting that we use only three stations in the leave-one-out situation, and have $k=3$ spatial parameters in the model, so the coefficients are poorly estimated. Therefore, capitalizing on low-cost sensors improves bias modeling and allows us to consider even more spatial covariates.

\section{\textcolor{black}{Discussion}}

This work proposes an innovative way to combine information from air quality physico-chemical models and multi-source measurements. More specifically, we make use of deterministic model outputs, which are spatially dense, and two types of in-situ measurements, one that is spatially sparse but more accurate, and the other that requires calibration but is cheaper and therefore denser. We propose a model that links both deterministic model outputs and measurements, and estimate its parameters through a Bayesian framework using \textcolor{black}{Markov chains Monte Carlo} methods. This Bayesian approach allows to adapt the model for the bias of deterministic model outputs to each specific city, and can serve as a tool for air quality monitoring agencies.

The study demonstrates that Bayesian inference can be used to correct both low-cost sensors measurements and model outputs. Indeed, SIRANE's output has been improved by 2 \unit\ in \textcolor{black}{root mean square error}(resp. 1 \unit\ in mean \textcolor{black}{absolute error}) during the study, which corresponds to a decrease of \textcolor{black}{more than 12\%} (resp. \textcolor{black}{9}\%) in errors at monitoring stations sites. Moreover, the use of low-cost sensors expands spatial coverage, which has proven to improve estimation of bias coefficients. With this method, low-cost sensors measurements can be corrected in-situ, enabling regular updates to correction models when necessary.

However, this study revealed some limitations. First, the time period was insufficient to evaluate the need for recalibration of low-cost sensors measurements and a potential seasonal change in bias coefficients. Second, the definition of the bias modeling could be broadened with additional variables, such as traffic measurements that were not available for this study. It would also be interesting to add more sensors, and compare the impact of the number of sensors (and their locations) on the concentration map correction. 

Other considerations include the possibility of incorporating mobile measurements, which are becoming increasingly common in air quality monitoring networks (see \cite{BerteroUrban2020} for a simulation study of air quality mapping using mobile devices, \cite{GressentDataFusion2020} for an application). The use of mobile sensors would make it possible to spatially densify the network and thus characterize the effect of spatial covariates more precisely at a very low cost. This will be the subject of a future work.

Finally, incorporating spatial and temporal correlations into the $\epsilon$ term associated with measurements and/or the bias definition would significantly enhance the modeling framework. One approach to achieve this is by computing an empirical variogram of the model errors, which can help identify an appropriate family of kernel functions. This would allow the introduction of a spatially dependent error term within the model, governed by its own set of parameters to be estimated. For a comprehensive treatment of such kriging techniques, the reader is referred to \cite{CressieStatistics1993}. Alternatively, the challenge of integrating measurements from sources with varying levels of quality can be addressed through multi-fidelity modeling, as introduced by \cite{KennedyPredicting2000} and is currently under study.


\bibliography{sn-bibliography}
\newpage

\appendix
\color{black}
\section{SIRANE outputs}\label{app:outputs}

\begin{figure}[htb]
    \centering
    \includegraphics[width=0.8\textwidth, trim={0.5cm 2.1cm 0 0},clip]{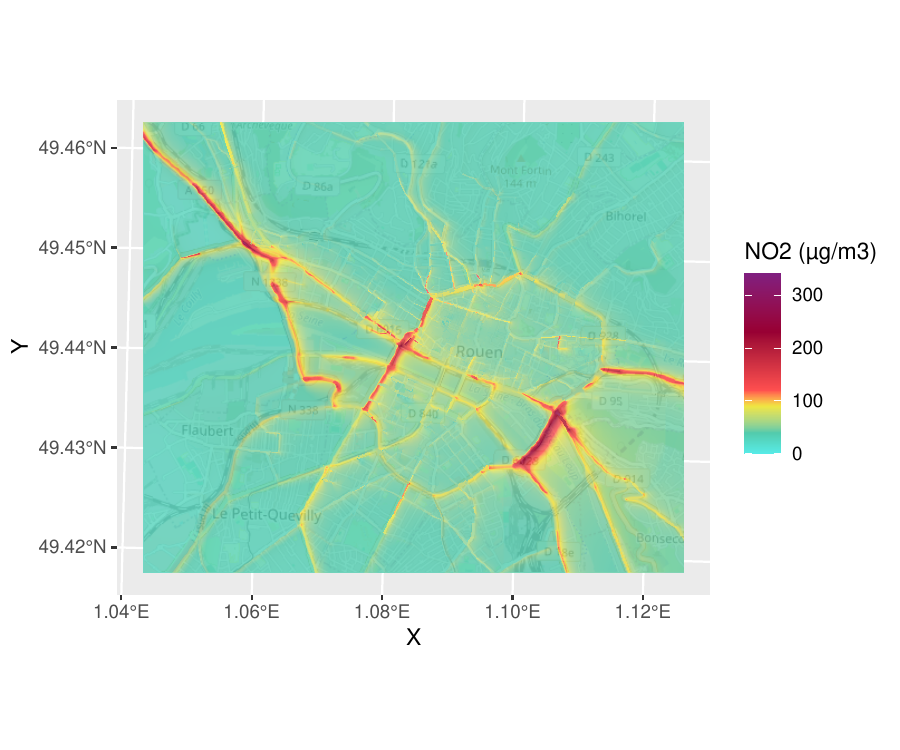}
    \caption{Example of SIRANE output for \no2 concentrations in Rouen on December 7, 2022, at 8:00 a.m.}
    \label{fig:ex-sirane}
\end{figure}

\color{black}
\section{Priors}
\label{app:priors}
We define the following prior distributions for the coefficients, that we assume to be mutually independent. We note: $\mathcal{N}(\mu,\sigma)$ the normal distribution with mean $\mu$ and standard deviation $\sigma$; $\Gamma(\alpha,\theta)$ the gamma distribution with shape $\alpha$ and scale $\theta$.
\begin{itemize}
    \item Bias:
        \begin{itemize}
            \item $a_0 \sim \mathcal{N}(0,1)$.
            \item $- a_c \sim ~ \Gamma(4,0.25)$
            \item $\theta_{T,1} \sim ~ \mathcal{N}(0,1)$ (with element 1 of $\theta_T$ standing for the $u_*$ coefficient)
            \item $- \theta_{T,2} \sim ~ \Gamma(2,0.001)$ (with element 2 of $\theta_T$ standing for the temperature coefficient)
            \item $\zeta_{T,1} \sim ~ \mathcal{N}(0.5,1)$ (with element 1 of $\zeta_T$ standing for the $u_*$ coefficient)
            \item $\zeta_{T,2} \sim ~ \Gamma(2,0.001)$ (with element 2 of $\zeta_T$ standing for the temperature coefficient)
            \item $- \zeta_{S,1} \sim ~ \Gamma(1,5)$ (with element 1 of $\zeta_S$ standing for the roads coefficient)
            \item $\zeta_{S,2} \sim ~ \Gamma(1,5)$ (with element 2 of $\zeta_S$ standing for the green spaces coefficient)
            \item $\zeta_{S,3} \sim ~ \mathcal{W}(2,1)$ (with element 3 of $\zeta_S$ standing for the elevation coefficient)
        \end{itemize}

    \item Sensors: $\forall j$,
        \begin{itemize}
            \item $\beta_j \sim \mathcal{N}(27000,1000)$
            \item $- \alpha_j \sim \Gamma(7,0.5)$
            \item $\sigma_j \sim \mathcal{W}(25,5)$
            \item $\gamma_{\text{NO},j} \sim \mathcal{N}(0,1)$
            \item $\gamma_{\text{Ox},j} \sim \mathcal{N}(0,1)$
            \item $\gamma_{\text{CO},j} \sim \mathcal{N}(0,1)$
            \item $\gamma_{\text{T},j} \sim \mathcal{N}(-10,5)$
            \item $\gamma_{\text{RH},j} \sim \mathcal{N}(-1,1)$
        \end{itemize}
\end{itemize}
with $\Gamma(k,\theta)$ being the Gamma distribution and $\mathcal{W}(k,\theta)$ being the Weibull distribution, both with shape $k$ and scale $\theta$.

\begin{table}[htb]
\centering
\begin{tabular}{|c|ccccccc|c|c|}
\hline
 Sensor & $\beta$ & $\alpha$ & $\gamma_\text{NO}$ & $\gamma_\text{CO}$ & $\gamma_\text{Ox}$ & $\gamma_\text{HR}$ & $\gamma_\text{T}$ & $\sigma$ \\ 
  \hline
  ASE4 & 26030.79 & -3.39 & 0.05 & -0.04 & 0.10 & -1.52 & -3.24 & 19.81\\ 
  ASE5 & 27028.88 & -3.18 & 0.01 & -0.01 & 0.08 & -2.04 & -6.92 & 20.07\\ 
  ASE6 & 28601.32 & -3.61 & 0.01 & -0.01 & 0.02 & -0.72 & -26.20 & 23.03\\ 
  ASE7 & 28210.28 & -3.60 & 0.02 & -0.01 & 0.03 & -1.94 & -3.13 & 12.25\\ 
  ASE8 & 23457.45 & -3.19 & -0.04 & 0.03 & 0.20 & -0.65 & -1.15 & 33.07\\ 
  ASE10 & 26533.46 & -3.69 & 0.03 & -0.01 & 0.07 & -0.77 & -8.78 & 21.44\\ 
  ASE11 & 28539.70 & -3.30 & 0.01 & -0.01 & 0.03 & -1.00 & -28.97 & 14.36\\ 
  ASE12 & 27471.70 & -3.46 & 0.01 & -0.02 & 0.08 & -1.53 & 1.62 & 19.59\\ 
  ASE13 & 26017.64 & -3.51 & 0.04 & -0.02 & 0.07 & -1.26 & -37.48 & 18.51\\ 
   \hline
\end{tabular}
\caption{Coefficients associated with the sensors during collocation period}
\label{tab:coloc_coefs}
\end{table}

We present in Table \ref{tab:coloc_coefs} the coefficients of sensor calibration linear model during a collocation period. Those results can be reproduced using available data detailed in \cite{DIB_Rouen2023}. As the percentage of explained variance (EV) of the linear model are of high quality (between 86 and 98 \%), the estimated coefficients might serve as base value for the distributions of $\alpha_j$, $\beta_j$, $\gamma_j$ and $\sigma_j$. However, the coefficients estimated with the proposed approach should be different, due to a change in sensors environment and sensors drifting in time.

\section{Corrected maps}
\label{app:correctedmaps}

\begin{figure}[H]
    \includegraphics[width = 0.6\textwidth, trim={0cm 0cm 1cm 0cm},clip]{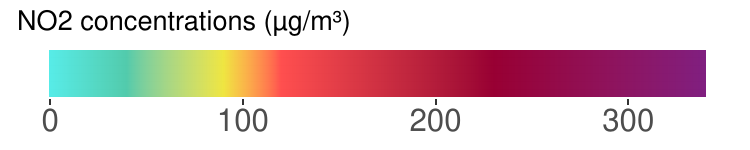}
    \centering
    \begin{subfigure}{0.48\textwidth}
        \includegraphics[width = 0.9\textwidth, trim={0.5cm 0.8cm 0cm 0.7cm},clip]{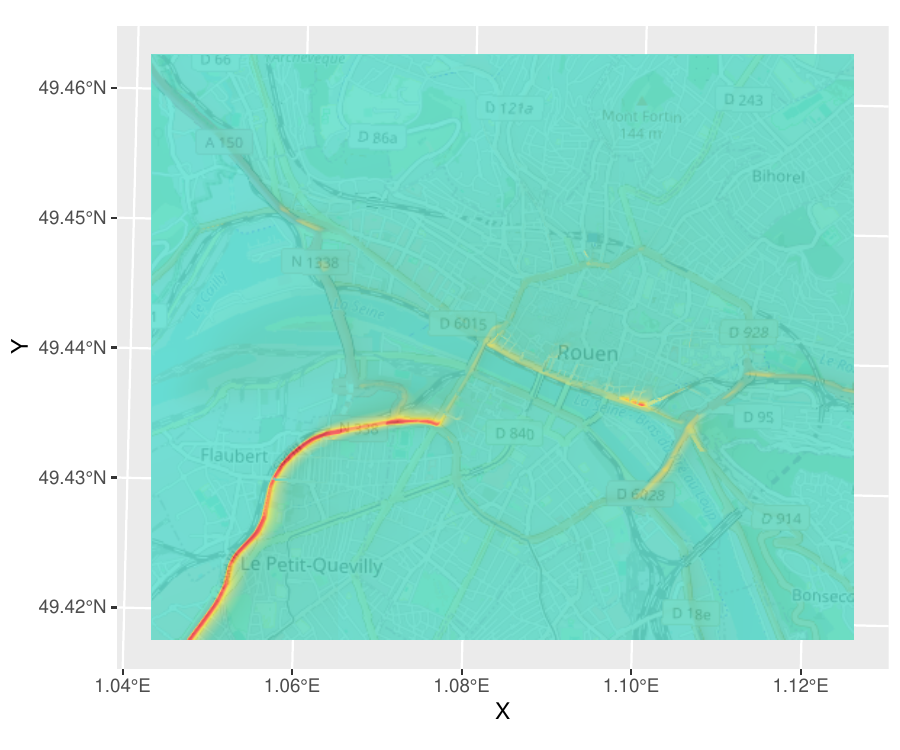}
        \caption{SIRANE output}
    \end{subfigure}
    \hfill
    \begin{subfigure}{0.48\textwidth}
        \includegraphics[width = 0.9\textwidth, trim={0.5cm 0.8cm 0cm 0.7cm},clip]{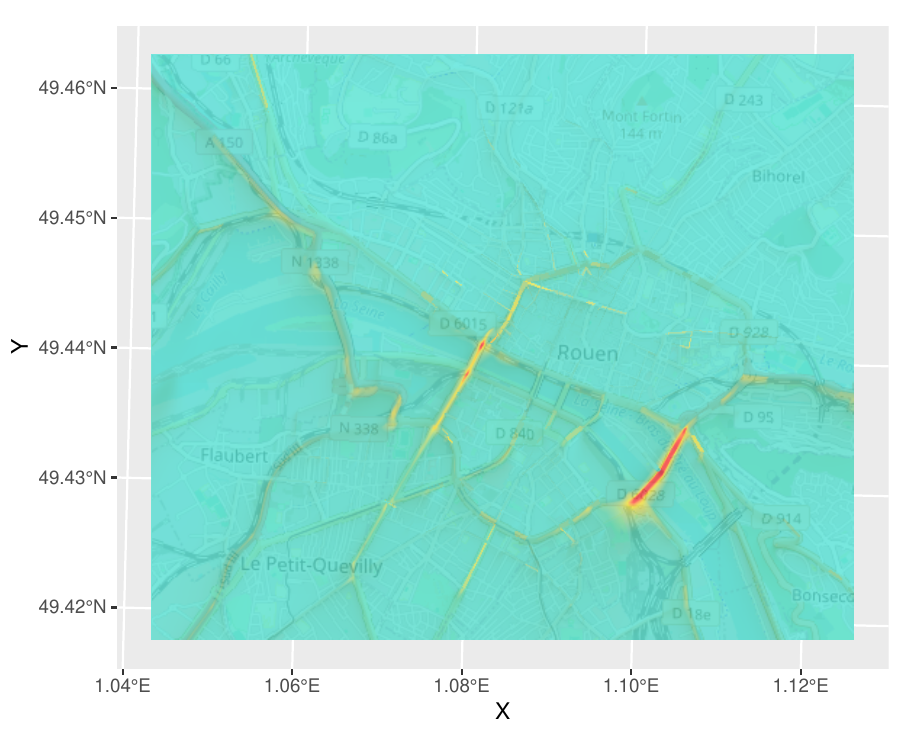}
        \caption{SIRANE output}
    \end{subfigure}
    \newline
    \begin{subfigure}{0.48\textwidth}
        \includegraphics[width = 0.9\textwidth, trim={0.5cm 0.8cm 0cm 0.7cm},clip]{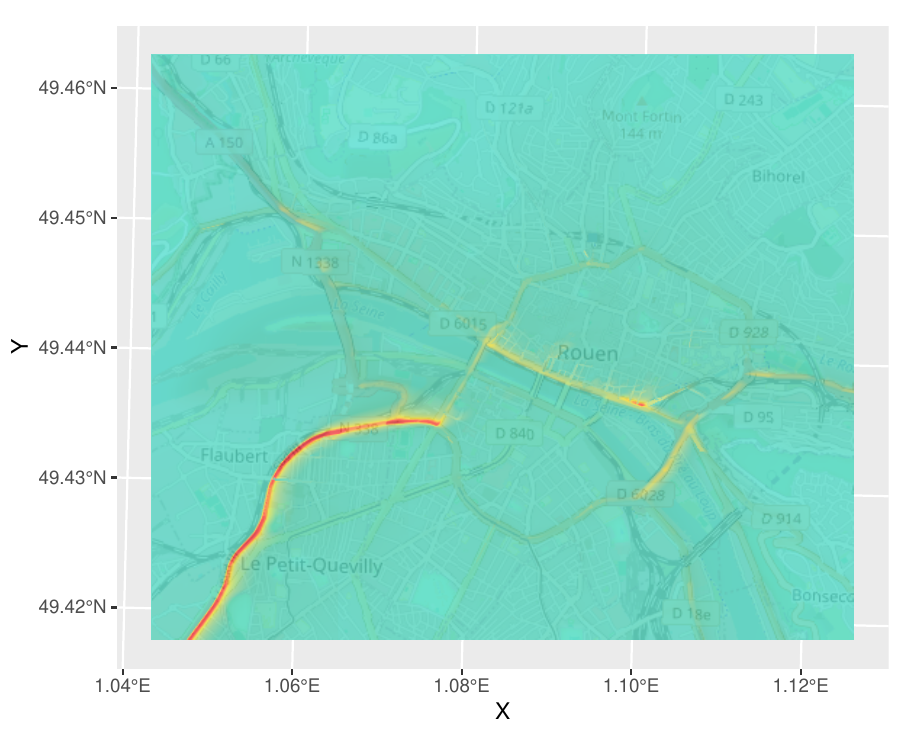}
        \caption{Correction using only stations}
    \end{subfigure}
    \hfill
    \begin{subfigure}{0.48\textwidth}
        \includegraphics[width = 0.9\textwidth, trim={0.5cm 0.8cm 0cm 0.7cm},clip]{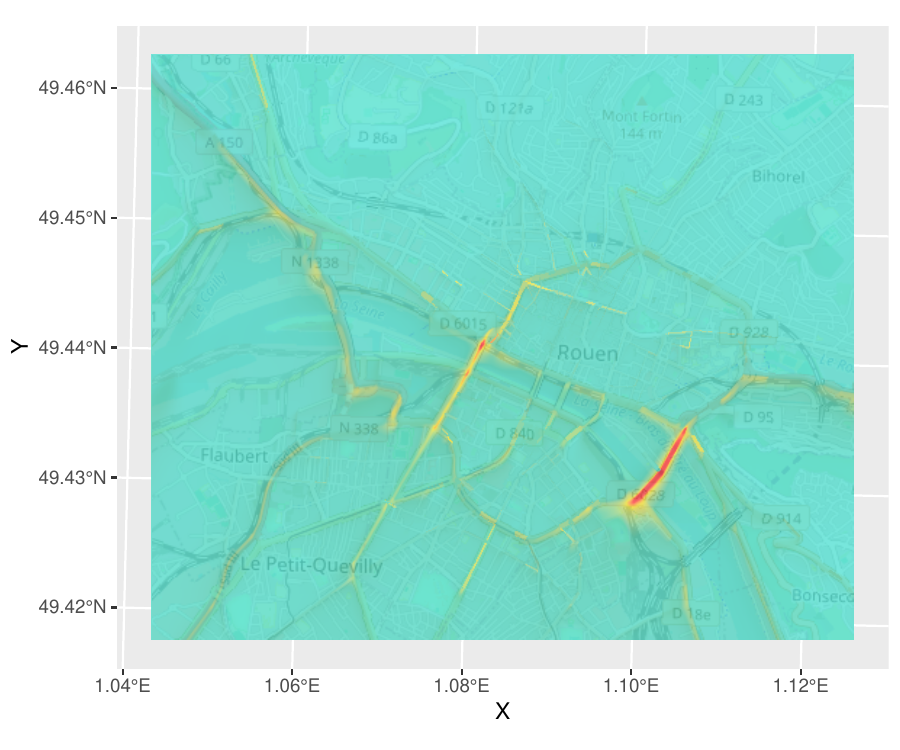}
        \caption{Correction using only stations}
    \end{subfigure}
    \newline
    \begin{subfigure}{0.48\textwidth}
        \includegraphics[width = 0.9\textwidth, trim={0.5cm 0.8cm 0cm 0.7cm},clip]{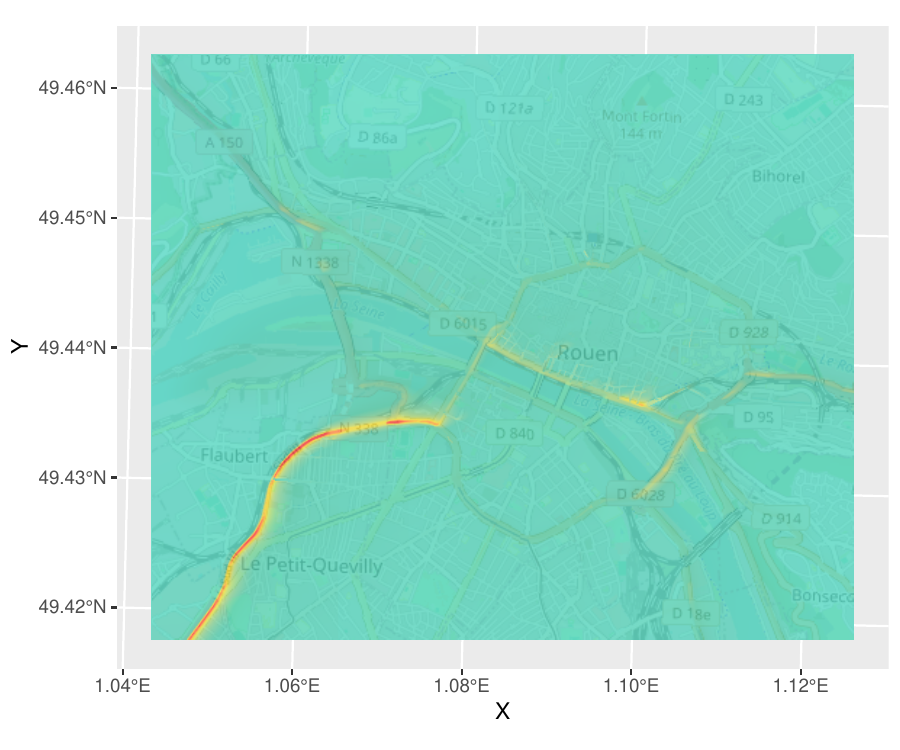}
        \caption{Correction using stations and \textcolor{black}{low-cost sensors}}
    \end{subfigure}
    \hfill
    \begin{subfigure}{0.48\textwidth}
        \includegraphics[width = 0.9\textwidth, trim={0.5cm 0.8cm 0cm 0.7cm},clip]{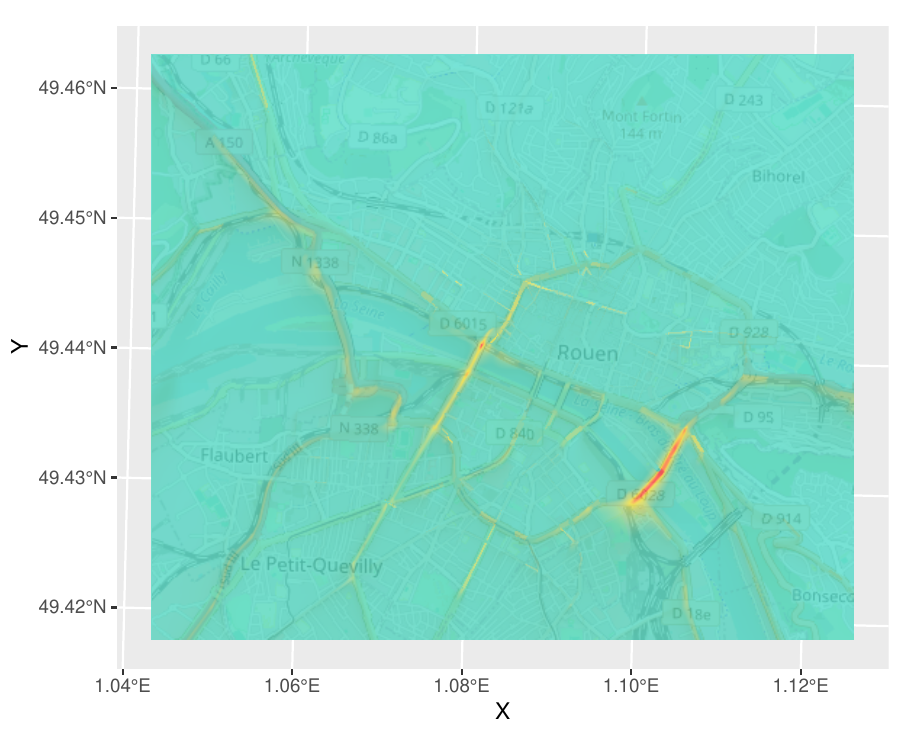}
        \caption{Correction using stations and \textcolor{black}{low-cost sensors}}
    \end{subfigure}
    \caption{Map of NO$_2$ concentrations in Rouen, on December 10, 2022 at 6 a.m. (left) and December 15, 2022 at 5 p.m. (right)}
    \label{fig:corrected-map}
\end{figure}

\textcolor{black}{Figure \ref{fig:corrected-map} presents two instants of the dataset that are of interest to illustrate the correction of SIRANE outputs. First, on December 10, 2022 at 6 a.m., SIRANE overestimated \no2 concentrations at reference stations (e.g. 104 \unit \ estimated at SUD3, while measurements were 45 \unit). These are shown in the left panel. Then, the figures on the right correspond to \no2 concentrations estimated on December 15, 2022 at 5 p.m. At that time, SIRANE underestimated concentrations on monitoring stations (e.g. 29 \unit \,estimated at SUD3, while measurements were 107 \unit). The colors on the map correspond to required color scale for \no2 concentrations in Europe based on the EAQI (European Air Quality Index).}

\textcolor{black}{One can notice that there are almost no differences between SIRANE's outputs and the maps corrected using only stations measurements. However, on December 10, 2022 at 6 a.m., the corrected map obtained using low-cost sensors appears slightly lighter, meaning that it has reduced estimated concentrations as intended. For example, at station SUD3, \no2\ concentrations measured were 45 \unit, while those estimated by SIRANE 104 \unit. The corrected values were 89 \unit. On December 15, 2022, at 5 p.m., the red zone on the east side of the city has been lightened, while the western zone (where station SUD3 is located) has been made redder. That is consistent with the \no2\ concentration at station SUD3 being underestimated by SIRANE at that hour (measurement of 107 \unit, estimation of 29 \unit, and corrected estimation of 36 \unit).}

\color{black}
\section{Diurnal variation}
\label{app:diurnalvar}
\begin{figure}[htb]
    \centering
    \includegraphics[width=0.8\linewidth]{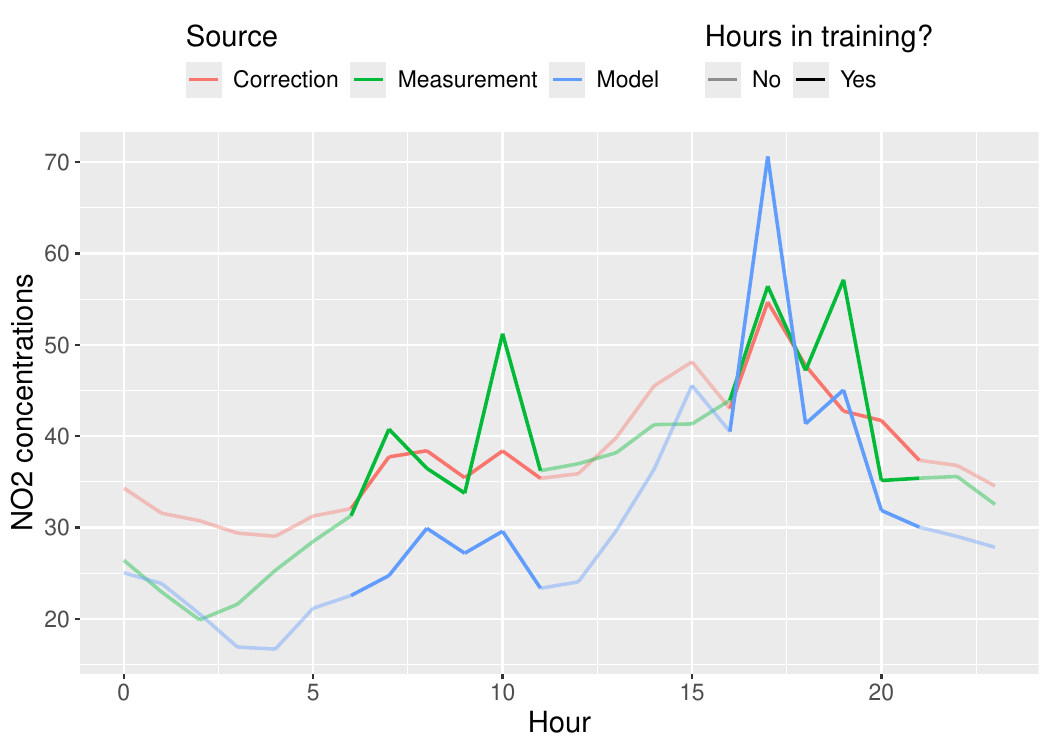}
    \caption{\color{black} Mean of measured and estimated \no2 concentrations depending on the hour of the day at reference station SUD3. The green line represents the measurements at the station, the blue one SIRANE estimations and the red one the corrected estimations. Means are computed excluding the training data. Opacity is higher if the hour has been considered in the training dataset.}
    \label{fig:sud3_hour}
\end{figure}

An aspect of interest is the respect of diurnal variation in the concentrations. As mentioned in Subsection \ref{subsec:validation}, \no2 concentrations are mostly produced by traffic emissions, and are thus more important around 8 a.m. and 5 p.m. Since hour is not a parameter in the correction procedure and is only accounted in the initial modeling, it seems interesting to look at the capacity of the correction model to reflect the changes in concentration by hour.
To illustrate this question, we corrected each SIRANE map that was not in the training dataset using both stations and low-cost sensors measurements.
Then, we computed the mean of measurements, SIRANE estimation and the corrected map estimations at each hour of the day and for each monitoring station. An example, based on the reference station SUD3, is given in Figure \ref{fig:sud3_hour}. 

In this figure, the hours of the day that were considered as the most polluted one are represented in full opacity. The others (during night and noon) are represented in lighter colours. It appears that the evolution of concentration in time are preserved with the correction, with higher concentrations estimated at 5 p.m., when the measurements are the most important. The correction is also closer to the measurement during daytime than the initial model. However, it can be noted that the correction does not perform well during nighttime, with greater estimated concentrations than the measured ones, which can be explained by the limitation of polluted hours in the training dataset.

\color{black}
\end{document}